\documentstyle[epsf,aps,twocolumn,prl]{revtex}
\begin{document}
\draft
\twocolumn[\hsize\textwidth\columnwidth\hsize\csname @twocolumnfalse\endcsname

\title{Sine-Gordon mean field theory of a Coulomb Gas}
\author{Alexandre Diehl, Marcia C. Barbosa, and Yan Levin}
\address{Instituto de F\'{\i}sica, Universidade Federal do Rio Grande do Sul \\
Caixa Postal 15051, 91501-970 \\
Porto Alegre, RS, Brazil}

\maketitle
\begin{abstract}

Sine-Gordon field theory is used to investigate the phase diagram of a neutral Coulomb gas. A variational 
mean field free energy is constructed and the corresponding phase diagrams in two (2d) and three 
dimensions (3d) are obtained. When analyzed in terms of chemical 
potential, the Sine-Gordon theory predicts the phase diagram topologically identical with the Monte Carlo 
simulations and a recently developed Debye-H$\ddot{\rm u}$ckel-Bjerrum (DHBj) theory. In 2d we find that the 
infinite order Kosterlitz-Thouless line terminates in a tricritical point, after which the metal-insulator 
transition becomes first order. However, when the transformation from chemical potential to the density is 
made the whole of the insulating phase is mapped onto zero density. 

\end{abstract}

]

The Coulomb gas provides a paradigm for the study of various models of critical phenomena\cite{Cardy}. 
In particular it is well known that the two dimensional (2d) Coulomb gas (CG) can be directly used to study 
the superfluidity transition in $^4$He films, arrays of Josephson junctions\cite{4HeJosep}, melting of 
two dimensional crystals\cite{Nelson80}, roughening transition\cite{Rough}, etc. Not withstanding its 
versatility our full understanding of the most basic model of Coulomb gas, namely an ensemble of hard 
spheres carrying either positive or negative charges at their center, is still lacking.

It is now well accepted that at low density the two dimensional plasma of equal number of positive and 
negative particles undergoes a Kosterlitz-Thouless (KT) metal insulator transition\cite{KT}. This 
transition is of an infinite order and is characterized by a diverging Debye screening length. Thus in 
the low temperature phase (insulator) all the positive and negative particles are associated into the 
dipolar pairs, while in the high temperature phase (conductor) there exists a finite fraction of 
unassociated, free, charges. As the density of particles increases the validity of the KT theory becomes 
questionable and the possibility of the KT transition being replaced by some kind of first order discontinuity 
has been speculated for a long time\cite{Berker}. The idea that there can exist a 
discontinuous transition between the insulating and conducting phases has gained further credence in view 
of the increasing computational power and an improving algorithm design needed for running large scale 
simulations of the particles interacting by a long range potentials\cite{Simulation}. Thus, it has 
been demonstrated quite convincingly that at high densities the KT infinite order line becomes unstable 
and is replaced by a first order coexistence between the low density insulating vapor and the high density 
conducting fluid-like phase. From the theoretical perspective, however, the nature of this metamorphosis 
is far from clear. At the moment there are appear to exist two competing views of what happens to the 
two dimensional plasma at higher densities. The first of this theories, presented by Minnhagen (Mh) 
et al.\cite{Minnhagen} in an series of papers going back ten years, predicts that the KT line will 
terminate in a {\it critical end point}, while the critical point of the coexistence curve separating the 
low and the high density phases lies in the conducting region. It is important to note that within the 
Minnhagen's theory the portion of the coexistence curve in between the critical point and the critical 
end point has both vapor (low density phase) and the liquid (high density phase) which are conducting. 
Although the Minnhagen's approach is often characterized as a version of Renormalization Group (RG), this 
is a misnomer. Its basis, which lies in a clever combination of a linear response formalism with some 
aspects of Sine-Gordon (SG) field theory, is much closer to the integral equations of liquid state 
theory than to the RG. The RG methodology being used more as a tool in studying the solutions of the 
integral equation found by Mh. 

\begin{figure}[h]
\vspace*{-4.3cm}
\begin{center}
\epsfxsize=6.cm
\leavevmode\epsfbox{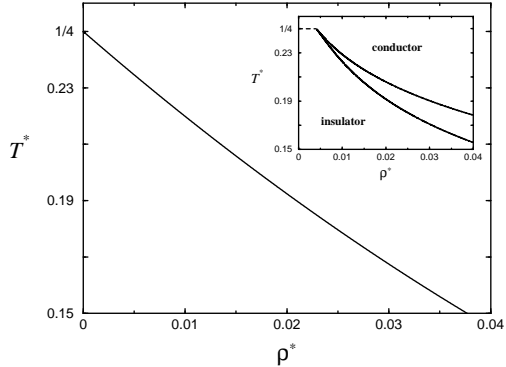}
\end{center}
\vspace*{.3cm}
\caption{Coexistence curves for the 2d SG field theory. The tricritical point is at $T^{\ast}_c\ =1/4$ and 
$\rho^{\ast}_c =0$. The inset represents the phase diagram for $d=2$, according to DHBj theory [9], 
with the KT line ending at a tricritical point (dashed line) localized at $\rho^{\ast}_c \simeq 0.003954$ 
and $T^{\ast}_c =1/4$. }
\label{coex2d}
\end{figure}

An alternative approach suggested by Levin et al.\cite{Levin94} is based on a recently developed 
Debye-H$\ddot{\rm u}$ckel-Bjerrum (DHBj) theory\cite{Fisher}. This method, which is intrinsically 
mean-field, relies on calculating the full electrostatic free energy of the ionic solution based on 
a linearized Poisson-Boltzmann equation. The effects of linearization are then corrected by allowing for 
the presence of dipolar pairs, the density of which is determined through the law of mass action. This 
theory has proven extremely powerful in elucidating the critical properties of a three dimensional 
(3d) electrolyte solutions\cite{Fisher}. In particular, the coexistence curve obtained on its basis was 
found to be in an excellent agreement with the recent MC simulations\cite{Panag3d}. The application of this theory to the 2d plasma has 
lead to a stark disagreement with the work of Mh. Where is Mh has found that the KT line terminates 
in a {\it critical end point} the DHBj theory predicts that it will terminate in a {\it tricritical} point 
after which the vapor {\it insulating} phase will coexist with a liquid {\it conducting} 
phase\cite{Levin94} (see inset Fig.\ref{coex2d}). 

Since the DHBj theory is intrinsically mean-field, one might argue that the fluctuations, such as a 
variation in dipolar sizes, might modify the phase diagram. This, however, is not very likely. It is 
well known that a properly constructed mean-field theory almost always retains the topology of the phase 
diagram upon inclusion of fluctuations. One of the few exceptions is when the volume of fluctuations is 
extremely large, such as in the case of transition between disordered and lamellar phases\cite{Brazovski} 
in magnets or amphiphilic systems. This, however, is not the case here. Furthermore the scaling of 
dipolar sizes can be included in a straight forward way into the DHBj theory, leaving the topology of 
the phase diagram unchanged\cite{XLi}. The metal-insulator line then becomes in exact agreement with the KT 
theory, in particular, giving the correct critical exponent $\nu =1/2$ for the divergence of the screening 
length upon the approach to the transition. The tricritical point persists, while the first order 
coexistence curve remains extremely narrow in the vicinity of the tricritical point. 

Comparing the predictions of Mh and DHBj theory to MC simulations we find that neither one is in a 
quantitative agreement with MC, which finds that the first order transition appears at a temperature which 
is significantly lower, and the density which is significantly higher than the prediction of either one of the above theories\cite{Simulation}. Nevertheless, the topology of the phase diagram observed on the basis 
of DHBj theory is the same as found in MC. Furthermore, the location of the tricritical point obtained in 
the MC corresponds closely to the region of the phase diagram where the narrow DHBj coexistence is found 
to swell significantly\cite{XLi}. 

The current impasse lead us to reexamine some of the foundations on which our understanding of CG is 
based. Most of the rigorous theorems concerning the nature of interactions inside the neutral plasma 
are based on the isomorphism between the CG and the Sine-Gordon Field Theory\cite[(b)]{4HeJosep}\cite{Edwards}. 
The mapping is exact only for the {\it point} Coulomb gas in the grand canonical 
ensemble. The short range repulsion is included post facto by introducing a suitable cutoff on all 
momentum space integrals. To what extend this procedure is valid is far from clear. Nevertheless, if the 
Sine-Gordon field theory is renormalized one obtains {\it exactly} the KT flow equations in terms of 
renormalized fugacity and temperature\cite{Knops}. This equations, however, remain valid only for low 
density (small fugacity). 

The attempts to construct a Sine-Gordon based mean-field theory go back to the work of Saito\cite{Saito79}, 
who has observed that already at the mean-field level the Sine-Gordon Hamiltonian knows about the metal 
insulator transition. In particular Saito was able to show that at low fugacity the Debye screening 
length diverged as $\xi_D = {\rm e}^{c/t^{\nu}}$, where $t=(T-T_c )/T_c$ and $\nu =1$. This should be 
compared with an equivalent expression obtained by KT but with $\nu =1/2$. Using field theoretic 
methodology Zhang et al.\cite{Zhang} extended the mean-field type of calculations of Saito and found that 
above a critical fugacity the screening length has a discontinuous jump from a finite to an infinite value. 
Zhang et al. then interpreted this point as a tricritical point terminating the continuous line of metal 
insulator transition. 

To compare the results of the Sine-Gordon based theory to the MC, one must be able to come up with a 
transformation from the fugacity, which is a natural variable in the field theoretical description, to 
the density, which is what the MC simulations measure. In the following we present a simple variational 
mean-field theory that accomplishes just that. It is in the process of transforming the phase diagram 
from the temperature-fugacity to temperature-density plane that the surprising new results were found. 

Our starting point is the Grand Canonical partition function for point particles of charge $\pm q$
\begin{eqnarray}
\label{gcpf}
{\cal Z} = \sum_{N_+ =0}^{\infty}\sum_{N_- =0}^{\infty}\frac{z_+^{N_+}}{N_+ !}
\frac{z_-^{N_-}}{N_- !}\;{\cal Q}(N_+ ,N_- )\;,
\end{eqnarray}
where
\begin{equation}
{\cal Q}(N_+ ,N_- )= \int \prod_{i=1}^{N}\frac{d^2 r_i}{\lambda^2} \exp \biggl[-\frac{\beta}{2D}\sum_{i\neq j}^{N}q_i q_j U(r_{ij})\biggr]\;.\nonumber
\end{equation}
Here $N=N_+ + N_-$ is the total number of particles immersed in a homogeneous medium of dielectric
constant $D$ and $\lambda =(h^2/2\pi mk_B T)^{1/2}$ is the thermal wavelength; the two-dimensional 
interaction term is $U(r_{ij})=\ln r_{ij}/a$, where $a$ is an arbitrary scale and $r_{ij}$ is the 
distance between particles $i$ and $j$. The fugacity is related to the chemical potential through 
$z_\pm ={\rm e}^{\beta \mu_\pm}$ and, along with the temperature ($\beta =1/k_B T$), determines all 
thermodynamic characteristics of the two-dimensional Coulomb-gas (2D CG).

To explore the thermodynamic properties of the above partition function it is convenient to map it onto 
the Sine-Gordon Field Theory\cite[(b)]{4HeJosep}\cite{Edwards}. Thus the partition function $\cal Z$ can 
be expressed as a functional integral over a real field $\phi$
\begin{equation}
\label{ZSG}
{\rm e}^{-\beta {\cal G}}\equiv {\cal Z}=\frac{\displaystyle \int {\cal D}\phi \;{\rm e}^{-H_{SG}}}{\displaystyle 
\int {\cal D}\phi \;{\rm e}^{-\int d^2 r \frac{1}{2}(\nabla \phi )^2}}\;, 
\end{equation}
where 
\begin{equation}
\label{heff}
H_{SG}=\int d^2 r \biggl[\frac{1}{2}(\nabla \phi )^2 - \frac{2{\bar z}}{a^2}\cos 
\biggl(\sqrt{\frac{2\pi \beta}{D}}q\phi \biggr) \biggr]
\end{equation}
is the effective Hamiltonian for a neutral Coulomb gas, and ${\bar z}=z(a/\lambda)^{2}{\rm e}^{\beta q^2 U(0)/2D}$ 
is the fugacity renormalized by a self-energy term. It is interesting to note that the saddle 
point of the Sine-Gordon Field Theory corresponds to the familiar Poisson-Boltzmann equation. In this 
paper, however, we will not use this analogy but instead construct a variational bound for the 
free energy. To this end we shall rely on the Gibbs-Bogoliubov-Feynman inequality exploring the convexity 
of free energy, ${\cal G}\leq G={\cal G}_0 + <H-H_0 >_0$, where ${\cal G}_0$ is the free energy 
associated with an arbitrary trial Hamiltonian $H_0$. The brackets indicate averaging over $H_0$. It 
is particularly convenient to choose as a trial Hamiltonian one having a Gaussian form, 
\begin{equation}
\label{h0}
H_0 = \int d^2 r \biggl[ \frac{1}{2}(\nabla \phi )^2 + \frac{m^2}{2a^2}\phi^2 \biggr]\;.
\end{equation}
In this case the free energy ${\cal G}_0$ and the average $<H-H_0 >_0$ are easily calculated, and we find 
\begin{eqnarray}
\label{freev}
\frac{\beta G}{V}= \frac{1}{8\pi a^2}\ln (1+m^2 ) - 
\frac{2{\bar z}}{a^2}\biggl(1+\frac{1}{m^2}\biggr)^{-1/4T^{\ast}}\;.
\end{eqnarray}
To perform the momentum space integrals the ultraviolet cutoff $(\Lambda =1/a)$, corresponding to the 
effect of the hardcore, was introduced\cite{cutoff}. The optimum upper bound is found by 
minimizing the free energy $G$ over all possible $m^2$. We find that the value of $m^2=m_0^2$ which 
leads to the optimal approximation to the real free energy $\cal G$ 
satisfies  
\begin{equation}
\label{minimiz}
m_0^2 =\frac{4\pi \bar z}{T^{\ast}}\biggl(\frac{m_0^2}{1+m_0^2}\biggr)^{1/4T^{\ast}}\;,
\end{equation}
where $T^{\ast} = k_B T D/q^2$ is the reduced temperature. The parameter $m_0^2$ is inversely proportional 
to the Debye screening length, $\xi_D$, inside the electrolyte solution, since it can easily be shown that 
the effective potential between two test particles separated by a distance $r$ is $V_{eff}(r)\sim <\phi (0)\phi (r)>_0$. 
Furthermore if $m_0^2=0\;(\xi_D =\infty )$ signifies that there is no screening which means that all the 
ions have paired up forming dipolar pairs. If this is the case the presence of an insulating phase 
is assured. 

\begin{figure}[h]
\vspace*{-4.4cm}
\begin{center}
\epsfxsize=6.cm
\leavevmode\epsfbox{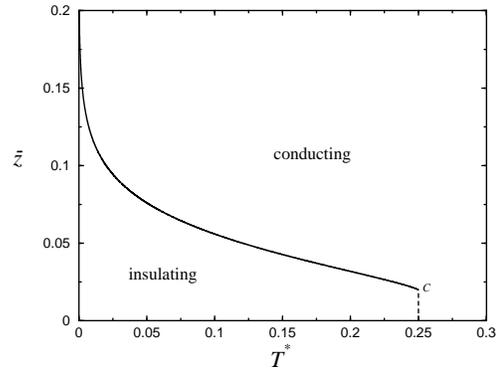}
\end{center}
\vspace*{.3cm}
\caption{Phase diagram of the 2D CG in the ${\bar z}-T^{\ast}$ plane. The solid line corresponds to the 
first-order transition, while the dashed line is the KT infinite order metal-insulator transition. The 
tricritical point $C$ is located at 
${\bar z}=1/16\pi$ and $T^{\ast}=1/4$.}
\label{z-T}
\end{figure}

Indeed from the equations (\ref{freev}) and (\ref{minimiz}) we find that the free energy possesses two 
minimas one of which is for $m_0^2 =0$ and the other $m_0^2 \neq 0$. The first order phase transition occurs 
when the free energies corresponding to the two local minimas become equal. The phase diagram in the 
$({\bar z}-T)$ plane for the 2D Coulomb Gas is presented in Fig.\ref{z-T}. It is essentially divided into 
two regions, each of which is characterized by a specific value of $m_0^2$: a conducting phase with a finite 
value of $m_0^2$ and an insulating phase with $m_0^2=0$. Separating these two phases there is a first-order 
transition line that ends at a tricritical point $C$ ($T^{\ast}_c=1/4$ and ${\bar z}_c=1/16\pi$). Below 
${\bar z}_c$ and close to $T_c^{\ast}$, the Eq.(\ref{minimiz}) can be approximated by 
\begin{equation}
\xi_D = \frac{a}{m_0}\sim a\;{\rm e}^{\displaystyle \scriptsize \frac{1}{2t^{\nu}}\biggl(\ln \frac{T^{\ast}}{4\pi \bar z}\biggr)}\;,
\end{equation} 
 
\vspace*{-.2cm}
\noindent where $t\equiv 1-1/4T^{\ast}$ and $\nu =1$. When $T^{\ast}\to 1/4$ the equation defines a 
line of a critical points $(\xi_D =\infty)$ that separates the conducting and insulating phases. 
This corresponds to the usual KT line of metal-insulator transitions. Contrary to the appearance, 
the first-order line and the critical line join smoothly at the tricritical point, with the tangency of the first-order line ensured by the divergence of $d\bar z/dT^{\ast}\simeq \ln |t|/4\pi$, when $t\to 0^-$.

As was emphasized in the introduction, in order to compare the results of our variational treatment with 
those of MC simulations it is essential to perform a transformation from the fugacity-temperature plane into 
the density-temperature domain. To this end we note that $\cal G$ is related to pressure and volume 
through ${\cal G}=-PV$, while the density is $\rho ={\bar z}\partial(\beta P)/\partial \bar z$. The 
transformation is then easily achieved and we find the coexistence curve presented 
in Fig.\ref{coex2d}. This curve is topologically identical to that obtained on the basis of pure 
linearized Debye-H$\ddot{\rm u}$ckel (DH) theory\cite{Levin94}. In particular we find that the high 
density conducting phase coexists with the {\it zero} density insulating phase. Namely, although the 
SG theory knows about the metal-insulator transition it can not give a proper account of the low 
density phase. Instead of producing dipoles the oppositely charged ions self annihilate on contact!

From our treatment it is not clear if this is a true property of the SG model or is an artifact of 
mean-field treatment, or maybe a result of the artificial way in which the hardcore was introduced into 
the model. Whichever the case, it is interesting to compare this result with the recently developed 
Debye-H$\ddot{\rm u}$ckel-Bjerrum theory (DHBj)\cite{Levin94} which, although also is of a mean-field type, 
does predict a finite density for the insulating phase. The coexistence curve for the DHBj theory is 
presented in the inset of Fig.\ref{coex2d}. 

The inability of SG model (at least at the mean-field) level to give a proper account of dipoles is also 
confirmed in $d=3$. In this case a first-order phase transition at low 
\vspace*{-.2cm}
\noindent temperatures between a low density (vapor) and a high density (liquid) phases is found as 
expected. Indeed, based on the DHBj theoretical\cite{Fisher} studies and simulations\cite{Panag3d}, the critical point is localized at $T^{\ast}_c \simeq 0.057$ and $\rho^{\ast}_c \simeq 0.025$. The resulting coexistence curve that emerges from our variational treatment predicts 
a critical point at $T^{\ast}_c \simeq 0.0565$ and 
$\rho^{\ast}_c \simeq 0.00135$. While the critical temperature $T^{\ast}_c$ is in agreement with previous results, 
the critical density $\rho^{\ast}_c$ is too small. This values should once again be compared 
with the pure linearized DH theory which does not account for the existence of dipoles; in that case it 
was found that $T^{\ast}_c =0.0625$ and $\rho^{\ast}_c =1/64\pi \simeq 0.005$. The underestimate of 
the critical density clearly indicates that at least at the mean-field level the SG theory, just as pure 
DH theory, does not give a proper account of non-linear effects such as the formation of dipoles.

How can these non-linearities be included is far from obvious. Why should the SG theory predict a metal-
insulator transition in the temperature-fugacity plane only to later map the whole of the insulating phase 
onto zero density? What is the proper class of diagrams that would have to be summed to produce a finite 
density for the insulating phase? This questions require serious attention if we wish to have a complete 
theory. Inability of the Sine-Gordon model, at least at the mean field level, to give a proper account of the low 
density phase might also be responsible for the distinct predictions between Mh and DHBj theories. In particular, it 
can be shown that the variational method that we have used corresponds to the summation to all orders of a 
certain class of diagrams. In the case of the standard scalar field theory this is the familiar Hartree-Fock 
approximation\cite{Parisi}. This class of diagrams is obviously insufficient if we are to believe that the 
SG theory can give a realistic account for the phase structure of the Coulomb gas. In his approach Minnhagen also 
relied on the SG theory to calculate the charge-charge correlation function. 
To this end he summed another set of diagrams. If that set was incomplete it could lead to some undesirable 
effects such as, for example, the wrong topology of the phase diagram. At the moment, however, this is only 
a speculation and a renewed theoretical effort is needed to study the Sine-Gordon field theory, now that it is 
evident that this model, besides the KT transition, also contains a first order discontinuity.


\begin{references}
\bibitem{cutoff} There is no unique relationship between the cutoff $\Lambda$ and the particle size. We 
adopt an isotropic sharp cutoff $\Lambda =1/a$. It should be remembered that the mapping of the 
Coulomb gas onto the Sine-Gordon field theory is only rigorously valid for point particles.
\end{references}
\end{document}